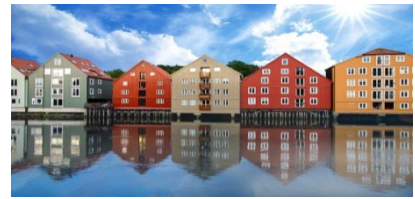

## Paper Information

## Summary

The urgent need to address climate change and pollution prompts societies worldwide to adopt carbon neutral energy and electrification. To facilitate this, a range of technologies and policies will be needed. Alternatives to traditional power grid reinforcement, such as grid-enhancing technologies and system automation, are particularly attractive due to their potentially low cost and fast deployment time. One of these alternatives is System Integrity Protection Schemes (SIPS) – automatic and curative remedial actions (RAs) which can boost grid transfer capacities without compromising with reliability since they can act faster than manual control. The use of SIPS however is scattered, with limited coordination between countries, and the full potential of using SIPS for capacity enhancement is not yet realized.

The aim of this paper is to provide a case study and comparison of SIPS usage in the Nordic countries, particularly in relation to capacity allocation. It also seeks to harmonize terminology relating to ancillary services, RAs, and SIPS. Finally, it examines and compares the inclusion of RAs and SIPS in different Capacity Calculation Methodologies (CCMs). In both dominant EU CCMs – Net Transfer Capacity (NTC) and Flow-Based (FB) – RAs play a pronounced role.

The paper is based on a survey and interviews with Nordic stakeholders, along with a literature review and analysis of public data. The main results indicate a large variation in SIPS use across the Nordics. Regarding terminology, we suggest that SIPS is a subcategory of RAs which overlaps with ancillary services. Concerning CCMs, NTC lacks the ability to fully capture the capacity constraints in meshed AC systems. This in turn makes it unviable to systematically enhance capacities using RAs. FB on the other hand explicitly includes RAs in the capacity domain. A lower bound for the economic value of RAs can be calculated, amounting to 11.5 million € in the Nordics in Nov and Dec 2024.

## Keywords

Automation, Capacity Allocation, Flow-Based, Protection Schemes, Remedial Actions, SIPS

# 1 Introduction

Globally, there are different naming conventions for power system automation intended to mitigate disturbances and stabilize operation under stressed conditions. *Remedial Action Schemes* (RAS), *System Integrity Protection Schemes* (SIPS), *System Protection Schemes* (SPS) and *Special Protection Schemes* (SPS) are all commonly used terms [1]-[2]. Moreover, the adjacent terms *remedial actions* and *ancillary services* are frequently used in EU grid codes [3], such as Capacity Allocation & Congestion Management (CACM) and the System Operation Guideline (SOGL). Finally, there is national legislation and TSO guidelines, adding an additional layer of protocols. Harmonization of terminology in this field is necessary, and a proposal for at least some new naming conventions and definitions will be laid forth in this section. It builds upon previous work in [1] and [2].

Among the terms above, we find *System Integrity Protection Schemes* (SIPS) the most adequate. This is the term used in IEEE Std. C37.250-2020 [4], recommended in [1], and in Cigré TB 742 [5] considered to encompass SPS, RAS, as well as underfrequency, undervoltage and out-of-step protection schemes. On a general level, SIPS consist of some input (a detection of disturbance or abnormal system condition), some output (a curative action mitigating the problem) and some logic in between. This is visualized in Figure 1, adapted from [6].

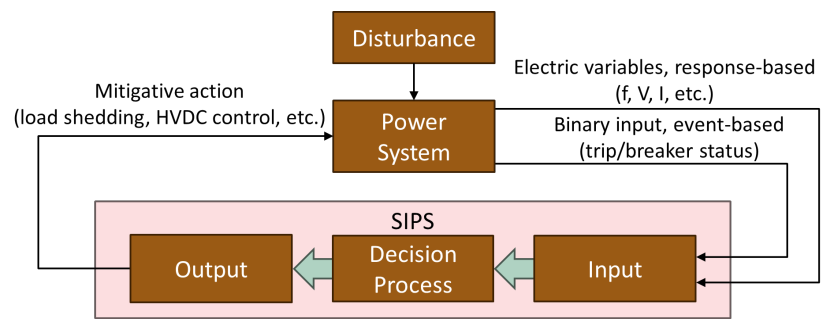

*Figure 1 – The general structure of SIPS in a power system, adapted from [6].*

SIPS can be used to counteract different kinds of critical system conditions, ranging from component overload to voltage and frequency instability. Mitigative actions can vary from grid reconfiguration and power control to generation rejection and load shedding [4]. The input is either a discrete event such as a trip (event-based) or a measured variable (response-based). Similarly, the output is either discrete, such as a trip of generation or load, or continuous, such as a ramp change of active or reactive power. Additionally, the geographic range of SIPS can vary from specific transfer corridors to entire synchronous areas [1]. SIPS differ from conventional protection in that the objective of SIPS is to protect the integrity of the power system and retain operational security, rather than to avoid damage of physical equipment. If SIPS counteract e.g. component overload, the purpose is primarily to avoid cascading outages.

## 1.1 Harmonization of terminology

*Remedial action* is defined in CACM, article 2(13) as "any measure applied by a TSO or several TSOs, manually or automatically, in order to maintain operational security". In SOGL, article 22 a range of examples of RAs are mentioned, including but not limited to tap-changing of transformers, modifying topologies, switching capacitors and reactors, requesting the change of reactive power output or voltage setpoint from synchronous generators or converters, redispatching, countertrading, adjusting active power flows through HVDC, curtailment, load-shedding and activating frequency management procedures.

*Ancillary service* on the other hand is defined in EU directive 2019/944 as "a service necessary for the operation of a transmission or distribution system, including balancing and non-frequency ancillary services, but not including congestion management". A similar formulation is found in e.g. Swedish national legislation [7]. A non-frequency ancillary service is then defined as "a service used by a TSO or DSO for steady state voltage control, fast reactive current injections, inertia for local grid stability, short-circuit current, black start capability and island operation capability".

Based on these definitions, there seems to be an overlap between RAs, ancillary services and SIPS. They all relate to operational security and can contribute to e.g. frequency stability or voltage stability. At the same time there appears to be some key differences: 1) the urgency of the problem, 2) who the agent is and 3) automation and range. We propose the following distinctions, visualized with examples in Figure 2:

- Remedial actions are only taken to mitigate detected, or avoid forecasted, disturbances/abnormalities, whereas ancillary services can always be provided.
- Ancillary services are provided by network users (generators, consumers), whereas remedial actions include also actions taken by the operator itself, in own facilities.
- Remedial actions can be preventive or curative, manual or automatic.
- SIPS are automatic and curative remedial actions.

This classification aligns with the prevalent definitions of remedial ("giving or intended as a remedy or cure") and service ("an act of being of assistance to someone"). Moreover, ancillary services, RAs and SIPS need not be kept mutually exclusive. It also aligns with the CRAC implementation guide [8], stating that SPS is "a remedial action consisting in an automatic device triggered after contingency."

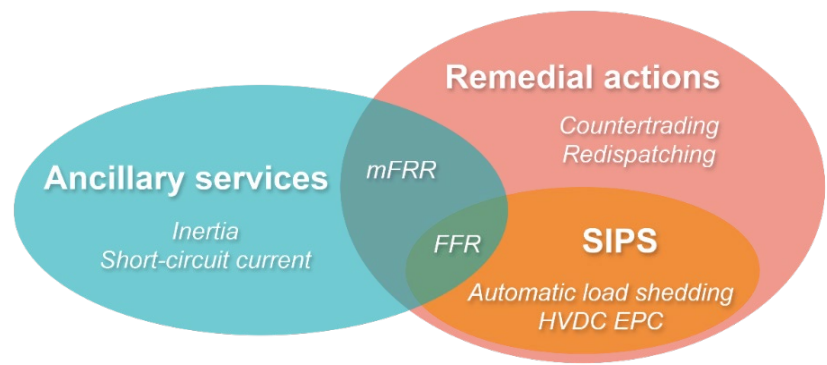

*Figure 2 – Proposed classification of ancillary services, remedial actions and SIPS with examples.*

The examples in Figure 2 illustrate how different resources fit into the different categories. Countertrading and redispatching for example are manual and preventive RAs, typically ordered to manage congestion, making them neither ancillary services nor SIPS. Automatic load shedding and HVDC Emergency Power Control (EPC) are hard-wired schemes, triggered by e.g. line trips or low system frequency, which renders them both SIPS and RAs. Inertia and short-circuit current are inherently provided and require no input or action, making them solely ancillary services. In contrast, FFR (Fast Frequency Response) acts as an alternative to inertia in low-inertia systems. Automatically triggered by low frequency, it delivers a burst of active power (typically from batteries) and thus qualifies in all categories. Finally, mFRR (manual Frequency Restoration Reserve) is manually triggered to restore normal frequencies, making it both an ancillary service (the provision) and a RA (the activation), but not a SIPS.

Note that the examples in Figure 2 do not provide a comprehensive list, and that there are dimensions not captured in the figure, for example in which system state the resource is provided or the owner of the facility. Examples are placed where they are likely to belong, but exceptions occur. Reactive power and voltage control can belong to several of the categories depending on the situation.

## 2 SIPS in the Nordics

SIPS usage varies across the Nordic countries in both implementation and purpose. As part of the SPS4SE project [9], a case study on SIPS in the Nordics was conducted, involving surveys and in-depth interviews with stakeholders along with a literature review. The stakeholder group involved representatives from all Nordic TSOs – Energinet (Denmark), Fingrid (Finland), Landsnet (Iceland), Statnett (Norway), and Svenska kraftnät (Sweden) – two Swedish subtransmission grid owners – E.ON and Vattenfall – as well as industry representatives from DNV and Hitachi. This section presents the main findings from the case study, a more detailed account can be found in [2]. Note that the case study was done before the harmonization of terminology presented in section 1, thus disparities might occur. Mitigative action is for example used in this context to specifically denote the output of SIPS, discerning it from RA.

### 2.1 Quantitative comparison

Quantifying the use or number of SIPS in a system is challenging due to their wide variation in range and complexity. In this study, we have used the definition that "all measurements and control actions used to mitigate the same instance of a critical system condition are considered to constitute the same SIPS". A critical system condition can for example be overload of a line section or transfer corridor, voltage instability in a region or frequency instability in the whole synchronous area. The results are presented in Table 1.

*Table 1 - Estimated number of SIPS in own system, on an interval scale.*

|  | Energinet | Fingrid | Landsnet | Statnett | Svenska kraftnät | E.ON | Vattenfall |
|---|---|---|---|---|---|---|---|
| # of SIPS | 10-100 | < 10 | 10-100 | > 100 | 10-100 | < 10 | < 10 |

From start, it became clear that Statnett employs SIPS to a larger extent than the other Nordic TSOs, also for capacity allocation (see section 4). A possible explanation is that the Norwegian grid is weaker and more decentralized than for example the Swedish and Finnish, increasing their relative benefit of using SIPS. Most of the SIPS at Statnett are generation rejection, load shedding or grid reconfiguration schemes to avoid congestion on specific lines or corridors.

Svenska kraftnät (Svk) also uses a notable number of SIPS, mainly generation rejection schemes to allow connection of additional generation. Energinet as well employs SIPS to increase capacities on specific sections. Fingrid in contrast is at the lower end of SIPS adoption among the Nordic TSOs, stating that their work on SIPS has only started.

## 2.2  Qualitative comparison

From a qualitative perspective, SIPS can be classified by their input-output combination, i.e. the critical system condition it targets and the mitigative action it triggers. All relevant stakeholders (TSOs and DSOs) were asked to identify combinations for which they had at least one SIPS. The results are summarized in Table 2. The critical system condition and mitigative action categories are adapted from [1] and [4] with minor adjustments.

*Table 2 - SIPS input-output combinations by operator, adapted from [2]. Actions are sorted by cost, conditions by criticality level or range. The four most common combinations are highlighted blue.*
*1 = Energinet, 2 = Fingrid, 3 = Landsnet, 4 = Statnett, 5 = Svenska kraftnät, 6 = E.ON, 7 = Vattenfall*

| Critical system condition | Mitigative action | | | | | |
|---|---|---|---|---|---|---|
| | Grid reconfig. | Var resched. / Q control | HVDC control | Generator / P control | Generation rejection | Load shedding |
| Component overload | 2, 3, 4, 5, 7 | | 1, 2 | 1, 3, 6 | 1, 4, 5, 6, 7 | 3, 4 |
| Abnormal voltage | | 5, 6, 7 | 5 | | | 3, 5 |
| Transient angle instability | 1, 3, 5 | 1 | 1 | 1, 3, 5 | 5 | 3 |
| Small-signal angle instability | 5 | | | 5 | | |
| Voltage instability | 1, 3, 5 | 1, 5 | 1, 2, 4, 5 | 1 | 6, 7 | 3 |
| Frequency instability | 1, 3 | | 2, 4, 5 | 3 | | All |

From Table 2, a few conclusions can be drawn. Four combinations are widely used, including grid reconfiguration or generation rejection addressing component overload, HVDC control addressing voltage instability and load shedding addressing frequency instability. Most combinations though are only ticked by one or two, indicating a variability in arrangements.

Svk exhibits the largest number of input-output combinations with 17 unique combinations. With four mitigative actions against abnormal voltage and four against voltage instability, the lack of voltage support in southern Sweden can be discerned. Svk is also the only operator with SIPS addressing small-signal angle instability.

## 3  Remedial actions in capacity calculations

In vertically disintegrated power systems, a central task for TSOs is to provide capacities to the market so that physical constraints are respected while the benefits of transmission are gained. In Europe, eight power exchanges handling the bulk power volumes traded have agreed on a common day-ahead (DA) market coupling principle, using an optimization algorithm called EUPHEMIA. The objective of EUPHEMIA is to maximize the total economic surplus for all regions, i.e. the sum of consumer surplus, producer surplus and congestion rent [10]:

To allocate capacities, which impose the optimization constraints, two main methodologies exist – NTC (sometimes ATC) and FB. In NTC, capacities are provided in each direction on each bidding zone border (BZB). The net position (NP) of a given BZ $A$ is allowed to vary within the sum of its inbound and outbound capacities according to [10]:

$$-\sum_{i=1}^{N} NTC_{B_i \to A} \leq NP_A \leq \sum_{i=1}^{N} NTC_{A \to B_i} \qquad (2)$$

where $B_i$ is a BZ bordering $A$, $N$ the number of BZs bordering $A$ and $NTC$ the allocated capacity for each respective border and direction. Internal constraints are either transposed to the BZB or neglected and dealt with later through countertrading/redispatching.

In FB, capacities are instead assigned through Critical Network Elements (CNEs) with Power Transfer Distribution Factors (PTDFs) and Remaining Available Margins (RAMs). This is a more granular approach, with the possibility to monitor individual lines, transformers and corridors, not just BZBs. In contrast to NTC, the FB approach with PTDFs captures how a change in NP of a BZ is physically distributed on all CNEs. The FB domain is defined as [11]:

$$\sum_{\forall n \in N} PTDF_{(c,n)} * NP_{(n)} \leq RAM_{(c)}, \qquad \forall\, c \in C \qquad (3)$$

where $N$ is the set of bidding zones, $C$ is the set of constraints in the flow-based domain, $PTDF$ are zone-to-slack power-transfer distribution factors and $RAM$ is the remaining available margin of a network element. The $RAM$ of a CNE is in turn defined as [12]:

$$RAM = F_{max} - F_{RM} - F_0 + F_{RA} + AMR - F_{AAC} - IVA \qquad (4)$$

where $F_{max}$ is the maximum allowed flow, $F_{RM}$ a reliability margin, $F_0$ the flow if all BZs operate at $NP = 0$, $F_{RA}$ an increase due to RAs, $AMR$ an adjustment to ensure $RAM > 0$, $F_{AAC}$ already allocated capacity for frequency reserves and $IVA$ an individual validation adjustment in case of unplanned outages or incorrect data. The $F_{RA}$-term ensures the explicit inclusion of RAs. A linearization is done around a two day-ahead (D-2) forecast, and the market coupling is performed using DC load flow. Note that a CNE can be critical in the basecase (N-0 situation) and under a contingency (a CNEC). The CNE can also be a combination of elements, such as a power transfer corridor (PTC) sensitive to voltage instability or oscillations [12].

In the CRAC implementation guide [8], it is stated how contingencies, RAs and additional constraints should be provided. This is essentially the same process in both NTC and FB, visualized in Figure 3. The capacity coordinator seeks the maximum flow that can be achieved on a BZB or CNE for a given set of contingencies, iteratively increasing the transfer by shifting generation and/or load on both sides. The most limiting contingency determines NTC for the BZB, or $F_{max}$ for the CNE. Non-costly RAs should be applied to increase the value.

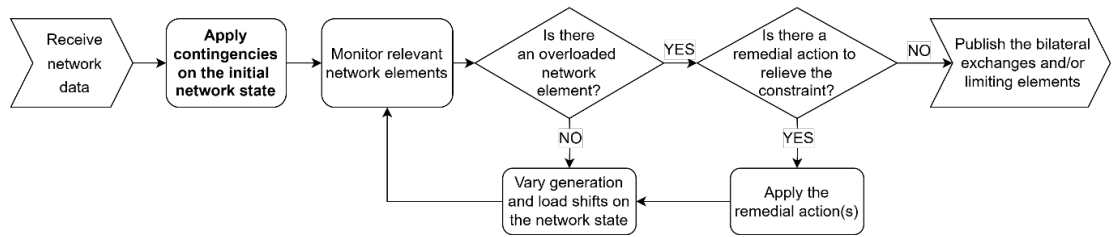

*Figure 3 – Coordinated capacity calculation process, adapted from [8].*

In systems with only radial interconnections between BZs – or HVDC links which are fully controllable – NTC and FB perform equally well, provided that the BZBs correspond to the actual bottlenecks [13]-[14]. Thus, in e.g. the South-West Europe (SWE) and the Baltic Capacity Calculation Region (CCR), consisting of topologically radial BZs, the NTC methodology is still used. The SWE CCR furthermore applies remedial action optimization to find the combination of RAs which gives the highest capacity across their borders [15].

In meshed AC systems however, such as the Core (central Europe) and the Nordic CCR, considerable loop flows exist and NTC can not accurately capture the constraints in the grid. The flows depend on the market outcome, and to account for all possible outcomes the feasible region must be unnecessarily restrictive. NTC in this case also lacks a unique solution, thus capacities are to some degree arbitrarily allocated and at the expense of each other. For a detailed explanation, see the three-node example in [14]. Since NTC does not accurately capture the constraints, it also becomes infeasible to systematically enhance capacities using RAs. The higher adequacy of FB had the Nordic CCR follow Core and switch from NTC to FB in Oct 2024. Since the go-live of FB in the Nordics, the FB domain has been publically available and the use of RAs in capacity calculations much more explicit and transparent [12].

## 4   Economic value of remedial actions in flow-based

In this section, a real example of the FB domain with post-coupling parameters for a certain hour (2024-11-06, 17:00-18:00 CET) is discussed. This specific hour rendered the highest DA price in the Nordics in Nov 2024: 550 €/MWh in DK1, as seen in Figure 4 (left). In addition, the carbon intensity of electricity production [16] is shown for the same hour in Figure 4 (right).

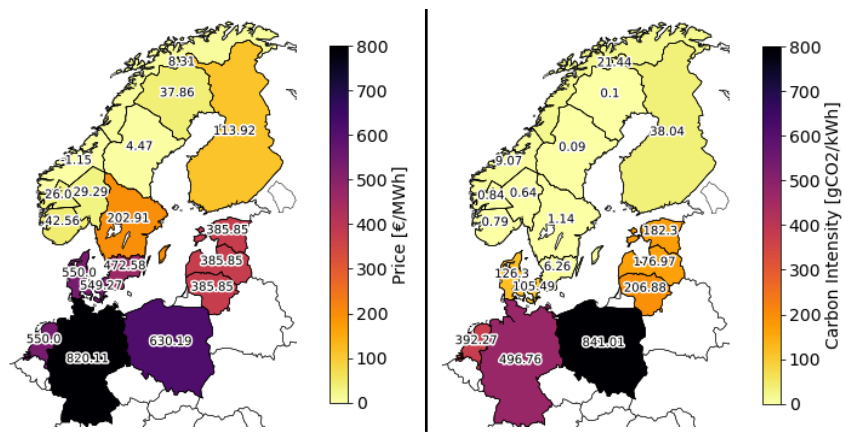

*Figure 4 – Left: Day-ahead electricity prices in Northern Europe [€/MWh]: 2024-11-06, 17:00-18:00 CET.*
*Right: Carbon intensity of electricity production [gCO2/kWh]: 2024-11-06, 17:00-18:00 CET.*

In Figure 5, two of the parameters in Equation (4) – $F_{max}$ and $F_{RA}$ – are displayed for all the active constraints (CNEs) at the time. $Fref$ represents the D-2 forecasted flow on the CNEs, which is the value the linearization is made around, whereas $FlowFb$ is the expected flow on the CNEs post-coupling. $MinFlow$ and $MaxFlow$ represent the boundaries within which $FlowFb$ is allowed to vary [12]. The CNEs are sorted by descending shadow price ($\lambda$), and all CNEs with a positive $\lambda$ are included. Shadow prices are an outcome of any constrained optimization and reflect the marginal value of relaxing the affiliated constraint with one unit. In FB, the shadow price of a CNE is the value of giving it an additional MW of capacity [14].

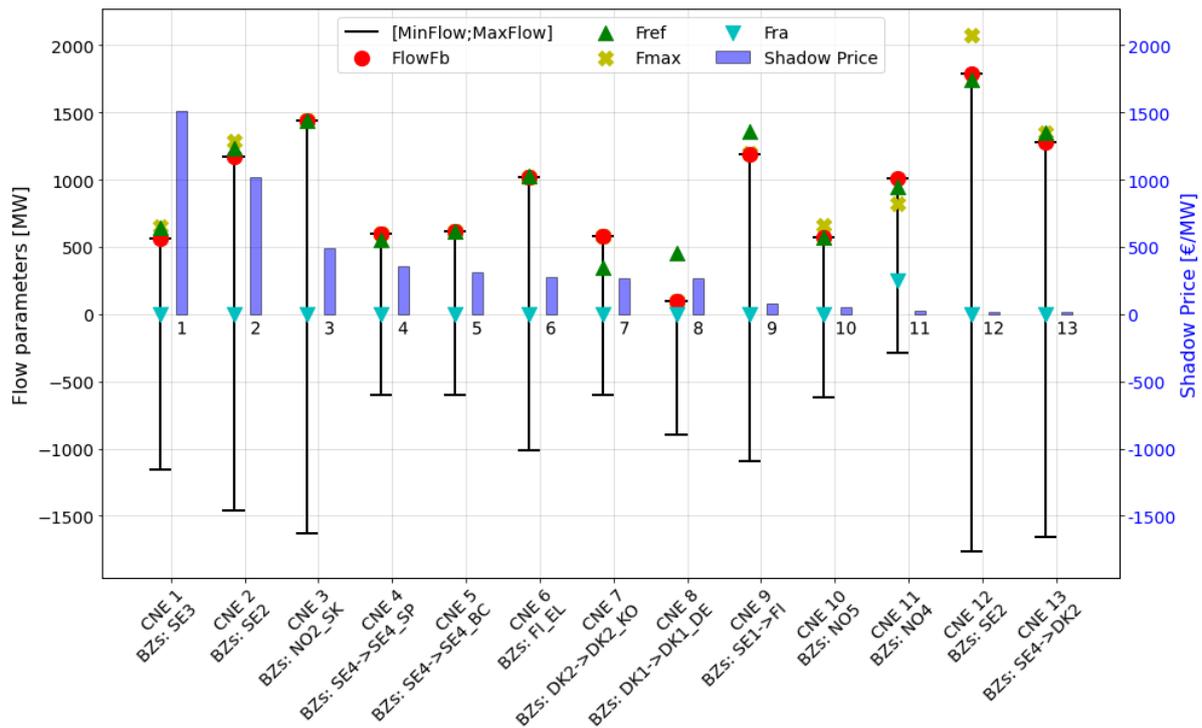

*Figure 5 – Flow-based domain with post-coupling parameters in the Nordics, 2024-11-06, 17:00-18:00 CET.*

The two most constraining CNEs in the Nordics during the hour shown in Figure 5 were internal CNECs in SE3 and SE2, with shadow prices of 1513 €/MW and 1019 €/MW respectively. $F_{RA}$ was zero for both, and additional capacity from SIPS or other enhancements would have provided significant value. CNE 11, located in NO4, had $F_{RA}$ = 250 MW. Here, RAs contributed with an economic value of at least $F_{RA} * \lambda$ = 250 MW * 29.3 €/MW = 7300 €. This value is a lower bound, since without the addition of $F_{RA}$ the constraint would have been tighter and $\lambda$ almost certainly higher. Additionally, some CNEs might have avoided becoming active constraints in the first place (i.e. $\lambda = 0$), due to the $F_{RA}$ contribution to $RAM$.

During the hour with the highest price in the Nordics in all of 2024 (2024-12-12, 17:00-18:00 CET), the second most constraining CNE was a PTC SE2 → SE3. At the time, this corridor from northern to southern Sweden had a shadow price of 722 €/MW, and $FlowFb$ was capped at 7494 MW. Considering the size and type of this constraint (PTC), this was likely a voltage or dynamic stability limit between SE2 and SE3, described in Annex 1 in [17].

In Figure 6 (left), the total $F_{RA}$ volume assigned on all CNEs in Nov and Dec 2024 is summed up and shown by TSO. This highlights that Statnett assigns much higher $F_{RA}$ volumes than the other Nordic TSOs. In Figure 6 (right), the cumulative sum of $F_{RA} * \lambda$ can be seen for the same period. This indicates that Statnett's use of RAs generated at least 10 million € in economic surplus value in the first two months. Increased capacity in and from the Nordics is likely to have climate benefits as well, enabling export of low-cost, fossil-free energy to displace marginal, fossil-based production (see Figure 4).

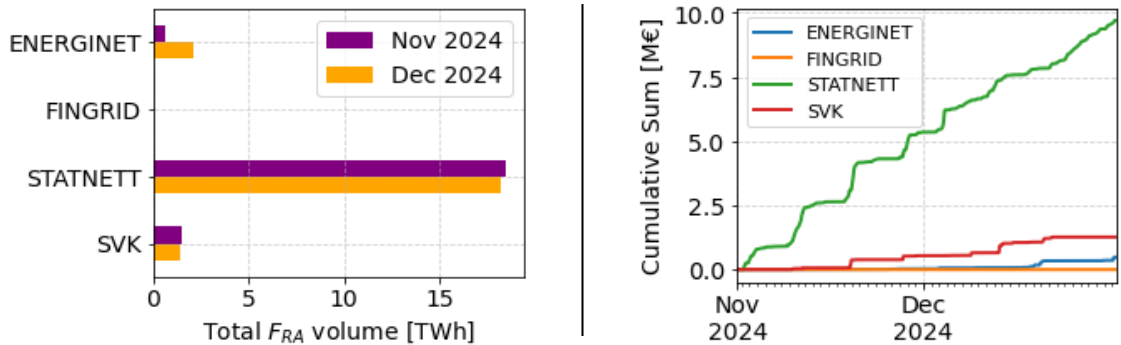

*Figure 6 – Left: Total $F_{RA}$ volume [TWh] by TSO in Nov and Dec 2024.*

*Right: Cumulative sum of $F_{RA}$ [MW] * Shadow Price [€/MW] by TSO in Nov and Dec 2024.*

## 5    Conclusions and discussion

In conclusion, there are different approaches to SIPS usage within the Nordic region. Quantitatively, Statnett displays the largest use and Fingrid the least. From a qualitative perspective, Svk has the largest variability in input-output combinations. We propose a harmonization of terminology where SIPS is a subcategory of RAs (automatic and curative) which overlaps with ancillary services. In meshed AC systems such as the Nordic, the NTC method is inadequate as it lacks the ability to adequately impose capacity constraints. This in turn makes it infeasible to systematically enhance cross-zonal capacities using RAs. If BZs are radially located, such as in South-West Europe and the Baltics, NTC becomes more viable and the use of RAs for capacity enhancement can be optimized. Since the go-live of FB in the Nordics, the inclusion of RAs in capacity allocation has become much more explicit and transparent, enabling calculation of e.g. RAs contribution to overall economic surplus.

Some caveats remain, however. It is for example not visible which type of RA increases $F_{RA}$, or if RAs are sometimes embedded in $F_{max}$. Furthermore, while FB is more suitable for the Nordic region than NTC, it still involves simplifications. Notably, FB requires linearization around a D-2 forecast, introducing both linearization and forecast errors. Additionally, FB is still a zonal model, aggregating large regions with estimated shift keys. Finally, since all $RAM$ parameters are expressed in MW and the market coupling relies on DC load flow, the ability to include non-thermal constraints such as voltage stability is limited and parallel simulations are required. Future studies should explore further how these non-thermal capacity constraints are calculated, and whether they can be alleviated using RAs and SIPS.

## Acknowledgments

This work is supported by the Swedish Energy Agency. The authors would like to extend their gratitude to the reference group members of the SPS4SE project.